\documentclass[12pt]{article}
\usepackage{natbib,graphicx,setspace,amsmath,amssymb,subfigure,url}
\newtheorem{prop}{Proposition}
\newtheorem{definition}{Definition}

\begin{document}

\title{Gaussian Process Models for Nonparametric Functional Regression with Functional Responses}         
\author{Heng Lian\\Division of Mathematical Sciences\\School of Physical and Mathematical Sciences\\Nanyang Technological University\\Singapore 637371\\Singapore\\E-mail: henglian@ntu.edu.sg}        
\date{}          
\maketitle
\begin{abstract} Recently nonparametric functional model with functional responses has been proposed within the functional reproducing kernel Hilbert spaces (fRKHS) framework. Motivated by its superior performance and also its limitations, we propose a Gaussian process model whose posterior mode coincide with the fRKHS estimator. The Bayesian approach has several advantages compared to its predecessor. Firstly, the multiple unknown parameters can be inferred together with the regression function in a unified framework. Secondly, as a Bayesian method, the statistical inferences are straightforward through the posterior distributions. We also use the predictive process models adapted from the spatial statistics literature to overcome the computational limitations, thus extending the applicability of this popular technique to a new problem. Modifications of predictive process models are nevertheless critical in our context to obtain valid inferences. The numerical results presented demonstrate the effectiveness of the modifications. 

\end{abstract} 

\section{Introduction}       
With recent ever-increasing interests and devoted efforts of many researchers, functional data analysis has developed into a full-fledged subarea of statistics. This surge of interests is explained by the ubiquitous examples of functional data that can be found in different application fields such as medicine, economics, physics etc., and many interesting applications can be found in \cite{ramsay02}. In such applications, use of specialized functional data analytic tools is preferable to using multivariate analysis on discretized finite-dimensional vectors, since with the former the smoothness property of the curves can be easily handled. With curves as the basic units of observation, analysis of functional data often provides interesting new statistical perspective and insights into modern data but also poses significant theoretical and practical challenges to the statisticians. 
 
  The literature contains a wide range of tools for functional data including exploratory functional principal component analysis, canonical correlation analysis, classification and regression. Two major approaches exist. The more traditional approach, carefully documented in the monograph \cite{ramsay05}, typically starts by representing functional data by an expansion with respect to a certain basis, and subsequent inferences are carried out on the coefficients. Another line of work by the French school \citep{ferraty06}, taking a nonparametric point of view, extends the traditional nonparametric techniques, most notably the kernel estimate, to the functional case. 

In this paper we are concerned with the functional regression problems with the extra complication that the dependent variables are also curves. Functional regression models with functional responses have been studied in \cite{ramsay05}, where two different linear models are proposed. More specifically, the integral model assumes that 
\begin{equation}\label{eqn:integral}
y(t)=\alpha(t)+\int_0^1 \beta(s,t)x(s)ds+\epsilon(t),
\end{equation}
where $\alpha(t)$ and $\beta(s,t)$ are unknown regression coefficients and $\epsilon(t)$ is the mean-zero noise. Note that we will assume throughout the paper that all functions are defined on the unit interval $[0,1]$ with trivial extension to functions defined on any bounded interval on the real line. On the other hand, the pointwise model assumes that
\begin{equation}\label{eqn:point}
y(t)=\alpha(t)+\beta(t)x(t)+\epsilon(t).
\end{equation}
Although the parameters reside in infinite-dimensional spaces for both types of linear models, in a functional data context both are usually deemed as parametric (see Definition 1.4 in \cite{ferraty06}). Recent works on parametric regression modeling include \cite{cuevas02,james02,cardot03,cardot05,muller05,jank06,hall07,crambes09,crainiceanu10}.

Nonparametric models for functional data only assumes a general regression form
\begin{equation}\label{eqn:nonpar}
y_i=F(x_i)+\epsilon_i, i=1,\ldots, n
\end{equation}
for a given independent and identically distributed (i.i.d.) sequence, where both $y_i$ and $x_i$ are functions. At least two approaches have appeared in the literature. The first method uses a simple kernel regression estimate \citep{ferraty04} and the second method is to use the reproducing kernel Hilbert spaces (RKHS) framework \citep{preda07}. 
Both methods were originally proposed for functional models with scalar responses. Studies of nonparametric methods for functional responses models are almost non-existent with the exception of \cite{lian07}, where both the kernel method and RKHS method are extended for functional responses. Although it is straightforward to extend the kernel regressor to this more complicated situation, extension of RKHS requires a novel concept, the so-called functional reproducing kernel Hilbert spaces (fRKHS). It was shown in \cite{lian07} that the fRKHS estimator performs better than the kernel estimator in functional response models. However, there are several limitations with the fRKHS framework:
\begin{enumerate}
\item The algorithm presented in \cite{lian07} only computes prediction estimates, but performing statistical inferences is difficult.
\item In \cite{lian07} the bandwidth parameters involved in the kernel functions are chosen based on heuristics leading to suboptimal estimates. 
\end{enumerate}
Although both problems could possibly be tackled using the frequentist approach, here we propose a fully Bayesian method to solve these problems. This is made possible by a connection between the fRKHS framework and the Gaussian process models. This type of connection is well-known in models with scalar responses \citep{craven79,wahba90} and we will present an extension in our context later. Besides, to overcome the computational limitations, we propose to adapt the predictive process models in the spatial statistics literature \citep{banerjee08} and also suggest two different modifications of the predictive process models to improve the prediction coverage accuracy. The advantages of working within the Bayesian framework are obvious - the unknown parameters and unknown regression function itself can be treated in a unified manner, and inferences about arbitrary unknowns are straightforward using the posterior distributions. As far as we know this is the first Bayesian study on nonparametric functional regression analysis with functional responses. Other works in Bayesian functional data analysis, including \cite{thompson08,rodriguez09,scarpa09}, do not deal with regression problems with a functional predictor. The studies carried out in \cite{shi07,shi08} are also quite different from ours, where the model is basically pointwise in nature and thus the predictors used are in fact multivariate instead of functional.

The rest of the paper is organized as follows. In Section 2 we briefly review the fRKHS framework. This section serves only as a motivation for our Bayesian proposal and the readers can choose to skip it without seriously hindering understanding of the later sections. In Section 3, after pointing out the connection of the fRKHS estimator with the Gaussian process models, our fully Bayesian approach is presented with a Markov chain Monte Carlo (MCMC) algorithm for posterior inferences. Furthermore, the predictive process models and their modifications are proposed as computationally more efficient approximations. In Section 4, simulation examples as well as a weather data application illustrate the advantages of our approaches. Currently the data size that we can deal with are limited to $nT\approx 5000$ where $n$ is the sample size and $T$ is the number of time points on which measurements are made on each function, simply because first of all we cannot store $nT\times nT$ matrices with larger dimensions (R will give an error message when attempting to do so). Using bigger computer systems can potentially solve this problem but this is outside the scope of the current paper. Finally we will conclude with some discussions in Section 5.

\section{Review of functional RKHS}
With scalar reponses $y$ in the nonparametric model (\ref{eqn:nonpar}), the well-known RKHS estimator for the nonparametric regression problem is obtained as the minimizer of the regularized loss
\begin{equation}\label{eqn:rkhs}
\sum_i\{y_i-F(x_i)\}^2+\lambda||F||_H^2,
\end{equation}
where $H$ is a RKHS with associated inner product $\langle\cdot,\cdot\rangle_H$ and induced norm $||\cdot||_H$. By definition, a RKHS $H$ is a Hilbert space of real-valued functions in which the point evaluation operator $L_x:H\rightarrow R$, $L_x(f)=f(x)$ is continuous. The Riesz representation theorem then implies the existence of a positive definite kernel $K(\cdot,\cdot)$ that satisfies the reproducing property $\langle K(x,\cdot),f\rangle_H=f(x)$. Solving the optimization problem (\ref{eqn:rkhs}) over an infinite-dimensional problem is made feasible by the representer theorem. Readers unfamiliar with these concepts can find more properties and discussions about RKHS in \cite{wahba90}.

In regression models with functional responses, we start by assuming that $y(\cdot)$ belongs to some RKHS $H$. For simplicity we assume $x(\cdot)$  also belongs to the same space $H$ although this is not necessary. Since $F$ now takes values in $H$, we need an extension of the concept of RKHS.
\begin{definition}\citep{lian07}
A functional RKHS $\mathcal{H}$ is a subset of $\{F:H\rightarrow H\}$. It is a Hilbert space, with inner product $\langle\cdot,\cdot\rangle_\mathcal{H}$, in which the point evaluation operator is a bounded linear operator, i.e., $L_x: F\rightarrow F(x)$ is a bounded operator from $\mathcal{H}$ to $H$ for any $x$.
\end{definition}
It can be shown from the definition that there exists a kernel associated with $\mathcal{H}$, and different from the traditional RKHS, this kernel is operator-valued. 

In practice, the observations for $y_i$ are made only on a grid $0\le t_1<t_2<\cdots<t_T\le 1$. Note we assume that all responses are observed on the same grid although different observation time points for different responses can be handled without much difficulty (although with much more complicated notations). Let $Y=(y_1(t_1),\ldots,y_1(t_T),y_2(t_1),\ldots,y_n(t_T))^T$ be the $n T$-dimensional vector of responses. Under some assumptions that simplify the derivation, \cite{lian07} shows that with the minimizer of (\ref{eqn:rkhs}) (here the RKHS $H$ that appears in (\ref{eqn:rkhs}) should be replaced by the fRKHS $\mathcal{H}$)), the fitted values $\hat{Y}$ can be represented as
\begin{equation}\label{eqn:fitted}
\hat{Y}=(A\otimes K)(A\otimes K+\lambda I)^{-1}Y,
\end{equation}
where $A=\{a(||x_i-x_j||)\}_{i,j=1}^n$, $K=\{k(|t_i-t_j|)\}_{i,j=1}^T$ are $n\times n$ and $T\times T$ matrices respectively. Here $a(\cdot)$ and $k(\cdot)$ are two positive definite functions and $||\cdot||$ is the $L_2$ norm. In \cite{lian07} as well as our current paper, we use the Gaussian kernel $a(t)=exp\{-t^2/\rho_1^2\}$, $k(t)=exp\{-t^2/\rho_2^2\}$.    The details can be found in \cite{lian07}. The choice of the kernel bandwidth parameters $\rho_1$, $\rho_2$ is critical for the performance but difficult to estimate. In \cite{lian07}, the author uses some simple heuristics and takes them to be the average distances between pairs of predictors and pairs of time points respectively. This heuristics-based choice produces encouraging numerical results but is in general only suboptimal.

\section{Gaussian process models and approximations}
\subsection{Gaussian process models}
In the previous section, the fitted response values with the fRKHS estimator are given by (\ref{eqn:fitted}). Consider now a zero-mean Gaussian process $W(x,t)$ indexed by $x\in H$ and $t\in[0,1]$ with covariance structure 
\begin{equation}\label{eqn:cov}
EW(x,t)W(x',t')=s^2a(||x-x'||)k(|t-t'|), s^2>0.
\end{equation} 
We assume a Gaussian process model for the observations
\begin{equation}\label{eqn:gp}
y_i(t_j)=W(x_i,t_j)+\epsilon_{ij},
\end{equation}
where $\epsilon_{ij}\sim N(0,\tau^2)$ are independent observation noises with variance $\tau^2$, which is often called the nugget effect in the spatial statistics literature where the Gaussian process is usually indexed by locations in $R^2$. The Gaussian process model is also popular in the machine learning community as a technique for nonparametric smoothing \citep{ras06}. Here the important difference from these works is that the process is indexed by an extra functional covariate, which makes it usable for functional data analysis, and unlike what is prevalent in the machine learning literature, we take a fully Bayesian approach here. The covariance structure (\ref{eqn:cov}) is directly motivated by the fRKHS estimator (as indicated in the following proposition), although the construction of a new kernel as the product of two other kernels is not novel in itself. We also acknowledge the possibility of using more general non-separable kernels. However, this problem is quite complicated although some recent progresses have been made and can be found in the statistical literature dealing with spatio-temporal data \citep{environ05,environ08,sjs10}. We only consider the separable kernel due to its simplicity as well as its connection to the fRKHS estimator.

The following proposition is an easy consequence of some well-known properties of the multivariate Gaussian distribution, and shows the connection between the Gaussian process model and the fRKHS estimator.
\begin{prop}
Assume the Gaussian process model stated as in (\ref{eqn:gp}). For any fixed $x\in H$, $t\in [0,1]$, let
\[\hat{W}(x,t)=E(W(x,t)|Y=(y_1(t_1),\ldots,y_n(t_T))^T,\theta),\]
which is the minimum variance unbiased linear estimate of $W(x,t)$ given $Y$.
We have
\[ \hat{W}=(A\otimes K)(A\otimes K+\lambda I)^{-1}Y\]
with $\lambda=\tau^2/s^2$, where $\hat{W}=(\hat{W}(x_1,t_1),\hat{W}(x_1,t_2),\ldots,\hat{W}(x_1,t_T),\ldots,\hat{W}(x_n,t_T) )^T$.
\end{prop}
The result says that the posterior mode (or equivalently posterior mean) of the Gaussian process model (conditional on $Y$ as well as parameters $\theta$) is exactly the same as the fRKHS estimator, when the smoothing parameter $\lambda$ is appropriately chosen. It is an extension of existing results for smoothing splines to the fRKHS context. Such a result is not trivial however without the concept of fRKHS in the first place. This observation suggests that we can estimate functional responses models in a fully Bayesian framework with the likelihood determined by the Gaussian process model (\ref{eqn:gp}), thus harvesting several advantages of Bayesian methods as mentioned in the introduction. In particular, joint posterior credible intervals would be difficult to obtain by other current methods.

With Bayesian models, we need to assign prior distributions to $\theta=(s^2,\tau^2,\rho_1,\rho_2)$. Our general principle in choosing prior distributions is to include weakly constraining prior information whenever possible. Although it is advisable to use subjective prior information if possible, our simulation results show that the data can usually provide enough information on these parameters. Details of prior specification are left to the next section when we discuss specific simulation examples.

Inferences in our model proceed by sampling from $p(\theta|Y)$. We use simple random walk Metropolis steps with normal proposals on the logarithmic scale of the parameters, with proposal variances tuned manually to achieve reasonable acceptance rates. Prediction with a new functional covariate $x$ at time $t$ is based on samples from $p(W(x,t)|\theta,Y)$, one for one with each sampled $\theta$.  Both steps involve the evaluation of Gaussian likelihood and thus the inversion of the covariance matrix for $Y$, which is equal to $s^2A\otimes K+\tau^2I$, a $nT\times nT$ matrix. Even if the sample size $n$ and the time points $T$ are both small, their product can be big enough to make the inversion, which has a complexity that is cubic in $nT$, a limitation of the approach. 

\subsection{Approximation with predictive process models}
To overcome the computational limitations, predictive process models are proposed to deal with large spatial data sets \citep{banerjee08}, and can be directly adapted for our purpose here. Even though the predictive process models are used intensively in the spatial statistics literature, we believe this represents their first application in functional response models. Consider sets $X^*=\{x^*_1,\ldots,x^*_m\}$ and $S^*=\{t^*_1,\ldots,t^*_q\}$ with $m\le n$ and $q\le T$, which are usually called the ``knots" and may and may not be a subset of the original covariates and time points. The predictive process is defined as the interpolant of the Gaussian process conditional on the process values at the knots, given by $\tilde{W}(x,t)=E(W(x,t)|W^*)$, where $W^*=\{W(x,t),x\in X^*, t\in S^*\}$. $\tilde{W}$ itself is a Gaussian process with a degenerate covariance structure, in the sense that the joint (multivariate normal) distribution associated with multiple covariates and time points is in general singular. More specifically, let $A_{,*}={{\{a(||x_i-x_j^*||)\}}_{i=1}^n}_{j=1}^m$ be the $n\times m$ kernel matrix on functional covariates, $K_{,*}={{\{k(|t_i-t_j^*|)\}}_{i=1}^T}_{j=1}^q$ be the $T\times q$ kernel matrix on time points, $A_{*,}$ and $K_{*,}$ denote their transposes, and $A_{**}={{\{a(||x_i^*-x_j^*||)\}}_{i=1}^m}_{j=1}^m$, $K_{**}={{\{k(|t_i^*-t_j^*|)\}}_{i=1}^q}_{j=1}^q$, then the covariance matrix of $Y$ in the predictive process model is 
\[s^2 [A_{,*}\otimes K_{,*}][A_{**}\otimes K_{**}]^{-1}[A_{*,}\otimes K_{*,}]+\tau^2I.\]
Determinant as well as inverse of such a matrix can be accomplished using Sherman-Woodbury-Morrison-type computations \citep{harville97} in terms of only $mq\times mq$ matrices. It was also shown in \cite{banerjee08} that the predictive process is the best approximation for the parent process in some sense.

For prediction under the predictive process, \cite{banerjee08} suggested to first sample $\theta$ from the posterior, then sample $W^*$ one for one with sampled $\theta$ values and finally use the model (\ref{eqn:gp}) with $W$ replaced by $\tilde{W}$ to sample the responses. We note however that given $\theta$ one can marginalize out $w^*$ resulting in a more efficient prediction procedure.
\subsection{Modifications of the predictive process model}
The approximation by the predictive process is shown to only cause a slight underestimation of the prediction uncertainties in \cite{banerjee08}. The underestimation is caused by the more restrictive covariance structure which leads to overconfident inferences under the posterior distribution. However, in our simulations, the empirical coverage of prediction intervals based on the predictive process is severely lower than the target value. The reason could be two-fold. Firstly, the sample size in our simulations is much smaller than that presented in \cite{banerjee08}, thus the effect of approximation is more noticeable. Secondly, the nature of the Gaussian process indexed by functional covariates is very different from that indexed by spatial locations. With the latter, the spatial domain is easily densely covered by a regular grid on the plane, while for the former, the covariates present in the data is more sparsely scattered in the space of possible covariate curves. A prediction to be made on a functional covariate far away from the training data should be very inaccurate whatever estimation method used. However, for the predictive process, once the values of the process on the knots are fixed, there is no uncertainty associated with $\tilde{W}$ whatsoever, leading to severe under-coverage of the prediction intervals. In the following we propose two ways to solve this problem, one based on a simple post-processing of the predictive variance produced by the predictive process, the other a modification of the predictive model itself. These two methods perform similarly in practice. 

We first present the explicit covariance structure to further understand the approximation involved in the predictive process. For the rest of this section, we assume that the parameters $\theta$ are known or we condition on them. For the parent process, given $x_{tst}$ on which the value of $y_{tst}(t_j)$ or $W(x_{tst},t_j), 1\le j\le T$ is to be predicted (for simplicity we assume that we aim to predict $y_{tst}$ at the same time points as in the training data), the joint distribution of $W(x_i,t_j), 1\le i\le n, 1\le j\le T$ and $W(x_{tst},t_j),1\le j\le T$ is given by the multivariate normal distribution
\begin{equation}\label{eqn:block}
p(\{W(x_i,t_j)\},\{W(x_{tst},t_j)\})=N\left(\mathbf{0},\left[\begin{array}{cc}
							\Sigma & \Sigma_{,tst}\\
							\Sigma_{tst,} & \Sigma_{tst,tst}
								\end{array}
						\right]\right)
\end{equation}
where $\Sigma=s^2A\otimes K$, $\Sigma_{,tst}=s^2A_{,tst}\otimes K$, $\Sigma_{tst,}$ is its transpose and $\Sigma_{tst,tst}=s^2A_{tst,tst}K$ and we recall from the previous subsection that $A_{,tst}=(a(||x_1-x_{tst}||),\ldots,a(||x_n-x_{tst}||))$, for example. In the predictive process, we effectively replace $\Sigma$ by $\tilde{\Sigma}=s^2[A_{,*}\otimes K_{,*}][A_{**}\otimes K_{**}]^{-1}[A_{*,}\otimes K_{*,}]$, and similarly for other blocks in the covariance matrix. The predictive distribution for $\tilde{W}(x_{tst},t_j)$ conditional on $Y$ in the predictive process is thus
\[ q(\{\tilde{W}(x_{tst},t_j),1\le j\le T\}|Y)=N\left(\tilde{\Sigma}_{tst,}(\tilde{\Sigma}+\tau^2I)^{-1}Y, \;\;\tilde{\Sigma}_{tst,tst}-\tilde{\Sigma}_{tst,}(\tilde{\Sigma}+\tau^2I)^{-1}\tilde{\Sigma}_{,tst}\right)\]
(note that the inverse above is computed by the Sherman-Woodbury-Morrison formula during implementation). Since $a(t)\rightarrow 0$ as $t\rightarrow\infty$, the variance above is small if $x_{tst}$ is far away from all $\{x_i^*\}$, and we see mathematically the under-coverage effect. This is also counterintuitive: if the test point is far away from all training points, one would expect more uncertainties in prediction while the predictive process would be (incorrectly) overly confident. One simple solution would be to avoid approximating $\Sigma_{tst,tst}$ by $\tilde{\Sigma}_{tst,tst}$ in the lower right block in (\ref{eqn:block}), resulting in the predictive distribution
\[N\left(\tilde{\Sigma}_{tst,}(\tilde{\Sigma}+\tau^2I)^{-1}Y, \;\;{\Sigma}_{tst,tst}-\tilde{\Sigma}_{tst,}(\tilde{\Sigma}+\tau^2I)^{-1}\tilde{\Sigma}_{,tst}\right).\]
This is our first proposed modification method.

Our second modification consists of replacing all blocks in the covariance matrix of (\ref{eqn:block}), for example $\Sigma$, by the more accurate $\tilde{\Sigma}+\mbox{diag}(\Sigma-\tilde{\Sigma})$, where $\mbox{diag}(\cdot)$ is the diagonal matrix containing only the diagonal entries of the original matrix. This effectively uses $\tilde{\Sigma}$ as in the predictive process, but keep the diagonal entries from the parent process. Note that the Sherman-Woodbury-Morrison formula can still be applied in this case.  

The first modification proposed is slightly simpler than the second in that the only change made to the predictive process is at the final stage of prediction and in addition only the predictive variance is modified. However, conceptually, this modification actually results in an inconsistent Bayesian model, since the covariance structure for the training data and test data are different. The second method requires the entire inference procedure be modified, in particular the MCMC algorithm, but it is a valid Bayesian model using a special covariance structure. 

\subsection{Selection of knots}
Determining optimal knots is a challenging problem. Since our focus here is more on the Bayesian model itself as well as the predictive process approximation, we simply use a random subsample for $X^*$ and a regular grid on $[0,1]$ for $S^*$. For spatial data, \cite{stevens04} showed that a regular grid is more efficient than simple random sampling. However, it is unclear how to choose a regular grid in a function space, and thus random sampling is used for $X^*$. It was shown in \cite{banerjee08} that estimation is more sensitive to the number of knots than the design of knots locations.

Finally, a larger number of knots is certainly preferable, but this should be weighted against the computation resources available. It is advisable that one repeat the analysis with different numbers of knots, and compare the predictive inferences. Of course the range to search also depends on the computational resources. 

\section{Numerical Examples}
\subsection{Simulation from Gaussian process model}
The data are simulated from the Gaussian process model (\ref{eqn:gp}). We set the sample size to be $n=30$ and an equi-spaced grid of $T=40$ points on $[0,1]$. Each functional covariate $x_i$ is generated as a standard Brownian motion multiplied by $5$ with a random starting position uniform in $[0,5]$. Other parameters in the model are set to be $\rho_1=20, \rho_2=0.2, s^2=2, \tau^2=0.05$. A separate test set with sample size $200$ is used for validation, where we also test on the same regular grid of $40$ time points on the functional responses. For the prior distribution, we use the weakly informative inverse Gamma distributions $IG(2,3)$ and $IG(2,0.1)$ for $s^2$ and the nugget effect $\tau^2$ respectively. Since \cite{lian07} demonstrated that setting $\hat{\rho}_1$ to be the mean of all $||x_i-x_j||, i,j=1,\ldots,n$ and similarly setting $\hat{\rho}_2$ to be the mean of $|t_i-t_j|$ produces promising results, we use the data-informed weak prior that is uniform on $[\hat{\rho}_1/10,\hat{\rho}_1\times 10]$ and $[\hat{\rho}_2/10,\hat{\rho}_2\times 10]$ respectively. These priors give sufficient support on a wide range of values for the bandwidth parameters.

The analysis is carried out using the full Gaussian process model, as well as predictive process model with $m=30, q=10$ and $m=20, q=40$, together with the two modifications proposed. That is, we only consider knots for the functional predictors or knots for the time points, but not both. 

Our algorithm is implemented in R. For each simulation, four MCMC chains are run for 5000 iterations, and standard convergence diagnostics provided in the coda package show that one can use the first 1000 iterations as the burn-in period. The parameter estimates as well as their Bayesian 95\% credible intervals are shown in Table \ref{tab:gp}. The MSE reported is the prediction error of the Gaussian process $W$ on the test data based on posterior mean. Also reported is the empirical coverage of the 95\% prediction intervals as well as their average lengths. Note these intervals are pointwise in nature. Using the first modification, the parameter estimates as well as the predictive mean are the same as the predictive process model without modification, and thus omitted from the table.

Several observations can be made regarding the results reported in Table \ref{tab:gp}. Firstly, setting $m=30, q=10$ produced better estimates than $m=20, q=40$, suggesting the functional predictor plays a more important role in estimation. The performance of the predictive process model with $m=30, q=10$ is similar to the full model. Secondly, there is an unacceptable under-coverage for predictive process models without modifications. Thirdly, the predictive process model without modification (or with the first kind of modification) tend to overestimate $s^2$, which is expected since the overly restrictive model requires a larger $s^2$ to fit the variability seen in the data. Finally, in terms of estimation errors, empirical coverage rates and interval lengths, the two kinds of modifications produce similar results and there is no clear winner between the two.

In terms of computational time, the full model takes about 20 hours to produce the output involving $5000$ iterations on our HP workstation xw4400 with  Intel Core 2 Duo E6700 processor and 2GB memory running R in windows XP, while with $m=30, q=10$, the predictive model takes about 2 hours, with or without modifications.

\subsection{Simulation from functional regression model}
In this simulation, we return to our initial goal of estimation and prediction in functional regression models with functional responses. The simulated data are now generated from 
\begin{equation}\label{eqn:sim}
y_i(t_j)=\int_0^1 2\sin(2\pi (t_j-s))x^2(s)ds+\epsilon_{ij},
\end{equation}
with $x$ generated exactly as before and $\epsilon_{ij}\stackrel{i.i.d.}{\sim}N(0,\tau^2)$ with $\tau^2=0.2$. This model is nonlinear in $x$ and a linear functional regression model such as (\ref{eqn:integral}) would not work well. We again set the sample size to be $n=30$ and number of time points $T=40$, and another sample of size $200$ is generated as test data. Prior distributions are specified the same as in the previous simulation. Simulation results are presented in Table \ref{tab:fda}. 

Conclusions similar to those for the Gaussian process simulation can be drawn here based on the table and thus not repeated. Note that the Gaussian process model is used as a means to estimate the functional regression model, and thus we cannot talk about the true values of the parameters. However one can compare the parameter estimates of different approximations relatively to those produced by the full model. In particular, we see again that random sampling of the functional predictors leads to worse performances than choosing a small grid for the time points. Directly using the guide values $\hat{\rho}_1$ and $\hat{\rho}_2$ in the fRKHS estimator results in a worse prediction error of $1.428$, where cross-validation is used to choose the smoothing parameter. 

Next we compare the prediction errors of other competing methods. Table \ref{tab:comp} shows the prediction errors of different approaches on exactly the same data. Listed in the table are the results obtained using integral model (\ref{eqn:integral}) (LININT), pointwise model (\ref{eqn:point}) (LINPT), as well as the versions when using $x^2(t)$ in place of $x(t)$ in these linear models (LININT2 and LINPT2 respectively). Also shown are the nonparametric kernel estimator (KERNEL) \citep{ferraty04}, as well as an ``oracle" version of the kernel estimator (OKERNEL) where the noiseless functional responses (that is, using the responses from (\ref{eqn:sim}) without adding noises $\epsilon_{ij}$) are used. The smoothing parameters in all the linear models (penalizing the second derivatives of the coefficients), as well as the bandwidth parameters for the kernel methods (using Gaussian kernel), are obtained by searching over a fine grid and the parameters achieving the smallest prediction errors are used, thus giving some advantages to these methods. From the table, we see that except for the method LININT2, which performs best since the correct linear structure is specified, the nonparametric methods outperform the parametric linear methods. Using the simple kernel methods however produce much bigger prediction errors compared to using our Bayesian method. 

\begin{figure}
\centering\includegraphics[width=2.5in]{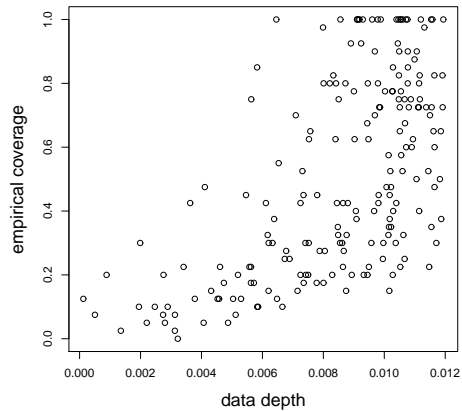}
\caption{Each of the $200$ test case is represented as a point showing its depth and empirical coverage rate on $40$ time points.\label{fig:depth}   }
\end{figure}

Since the functional regression model is really what we are interested in here, we take this opportunity to provide further insights into the extremely low empirical coverage rate of the predictive process model without modification. We have argued that the lower coverage rate might result from the fact that the predictive process models cannot correctly extrapolate to functional predictors far way from the majority of the covariates seen in the training data. For functional data, \cite{lopez09} introduced a novel notion of data depth that quantifies the outlyingness of any curve with respect to a reference set of curves. Roughly speaking, a curve that is a outlier compared to the reference curves has small depth. We compute the depth of the $200$ functional covariates in the test data with respect to the functional covariates in the training data. For each of the $200$ covariates, we also compute separately the empirical coverage rate based on the $T=40$ time points for each covariates. The relationship between the depth and the coverage rate is shown in Fig \ref{fig:depth}. A clear positive correlation is seen suggesting that the coverage rate is lower for new data farther away from the training data. 

\subsection{Illustration with the weather data}
\begin{figure}
\centering\includegraphics[width=5in,height=2.7in]{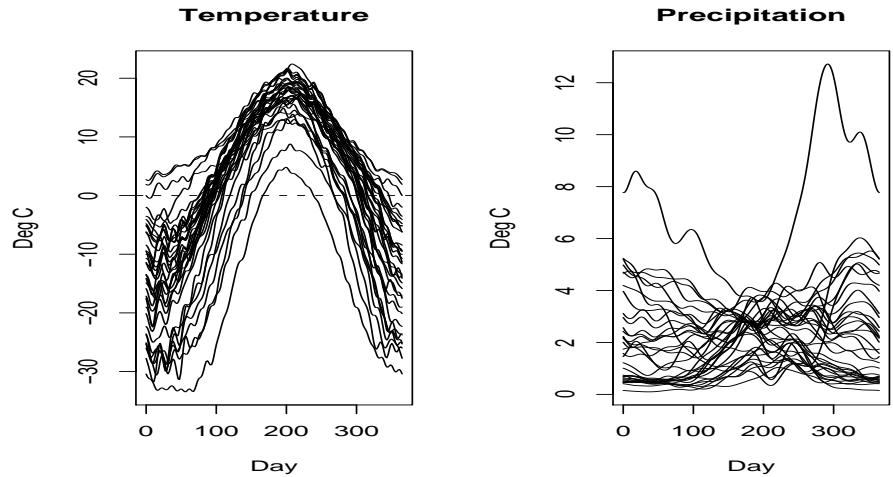}
\caption{Daily weather data for 35 Canadian stations. The curves plotted here result from using smoothing splines to fit the data.\label{fig:weatherdata}   }
\end{figure}

The daily weather data consist of daily temperature and precipitation measurements recorded in
35 Canadian weather stations. These are actually averages over the years from 1961 to 1994,
applying a correction for leap years. The smoothed data are plotted in Fig \ref{fig:weatherdata} (produced with
the sample code provided in the fda package). With $n=35$ and $T=365$, R cannot even allocation enough memory to store an $nT\times nT$ matrix on our workstation. Thus we subsample the data
and use only the weekly measurements (measurements from every 7th day), and then each observation
consists of functional data observed on an equi-spaced grid of 53 points. We treat the temperature
as the independent variable and the goal is to predict the corresponding precipitation curve given
the temperature measurements. As is previously done, we set the dependent variable to be the
log transformed precipitation measurements, and a small positive number is added to the values
with $0$ precipitation recorded before taking logarithm. The priors are set as in the simulation studies. We also tried several different priors. We use priors $IG(2,0.5), IG(2,3)$ or $IG(2,10)$ for $s^2$ and priors $IG(2,0.02), IG(2,0.1)$ or $IG(2,1)$ for $\tau^2$. The results obtained using these priors are almost identical. 
 
\begin{figure}
\centering\includegraphics[width=5in]{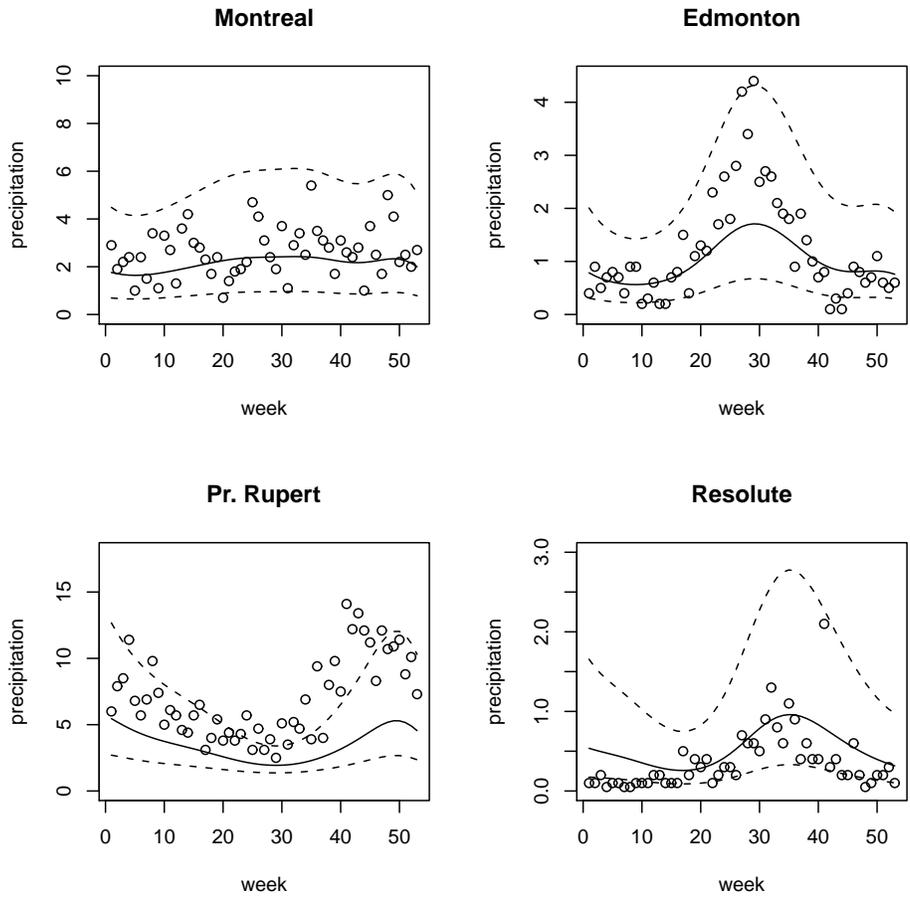}
\caption{Precipitation data (points), prediction estimate (solid) as well as pointwise credible intervals (dashed) using the full Gaussian process model. \label{fig:weatherfull}   }
\end{figure}

\begin{figure}
\centering\includegraphics[width=5in]{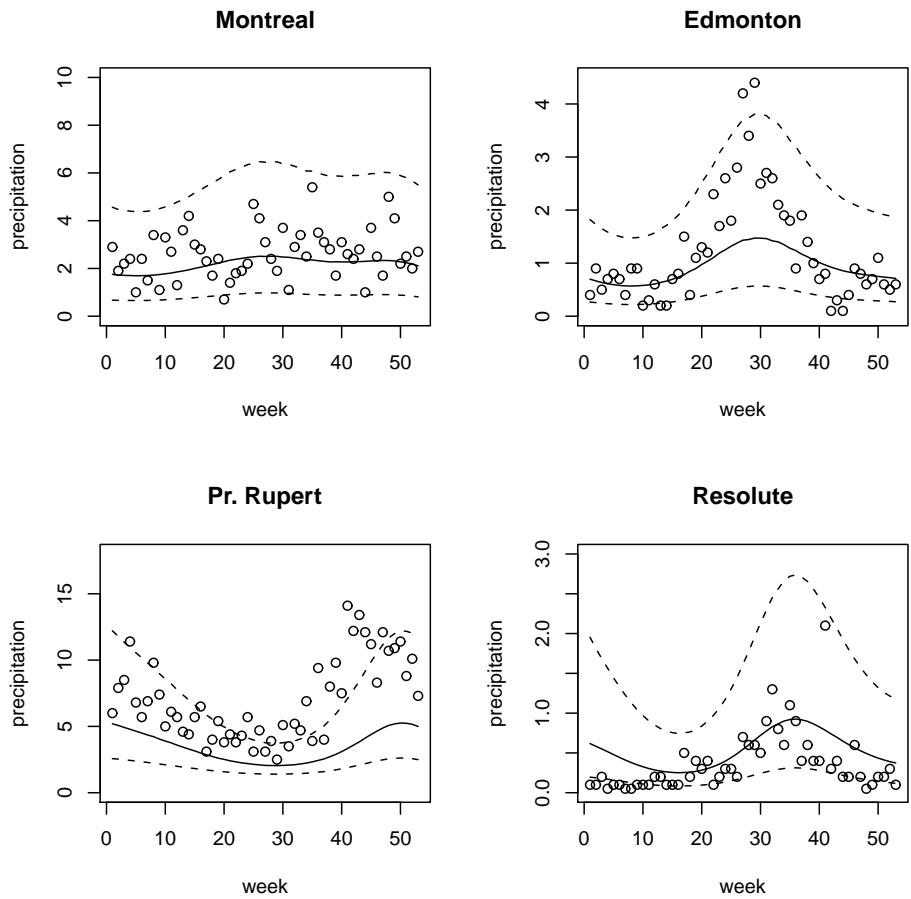}
\caption{Precipitation data (points), prediction estimate (solid) as well as pointwise credible intervals (dashed) using the predictive process model with the first kind of modification. \label{fig:weatherpp}   }
\end{figure}

  The prediction using the full Gaussian process model is shown in Fig \ref{fig:weatherfull} while that using prediction process models with the first kind of modification is shown in Fig \ref{fig:weatherpp}. These four stations are left out when training is performed on the rest. With much savings in computation for the latter, the predictions made by these two are very similar. One visual difference is that in the prediction for the Edmonton station, the peak in precipitation produced from the predictive process model is less conspicuous. Using the second kind of modification produces results almost identical to those shown in Fig \ref{fig:weatherpp}. The large error for Pr. Ruppert is due to the unusual wet weather there, which can also be visually identified as the outlier in the right panel of Fig \ref{fig:weatherdata}.

  For this dataset, the prediction errors on those four stations are compare to those obtained using the integral model (\ref{eqn:integral}) (Table \ref{tab:weather}). The Bayesian method outperforms the linear method for three of the four stations, with the most drastic improvement for the Pt. Ruppert station. This can be explained by the outlyingness of the responses for this station, since with outlying observations, the parametric model is severely misspecified while the nonparametric model is more robust with less assumptions imposed. The results obtained for the linear pointwise model are much worse and thus not shown here.

\section{Conclusion}
We have proposed in this paper a fully Bayesian approach to fitting a nonparametric functional regression model with functional responses. The Gaussian process model we have proposed is motivated by the theory of fRKHS but can also be constructed directly. We show the connection between the two and thus extend the existing knowledge on their relationship to the case of regression models with functional responses. To ease the computational burden, which depends on the product of the sample size and the number of time points and thus more of a problem here than functional regression with scalar responses, we adapt the predictive process models for our purpose and also propose two modifications to address the problem of under-coverage for the prediction intervals. We note that the under-coverage problem is particularly severe here compared to the previous findings in spatial data analysis, and we provided some insights into this problem.

We have tried using a Mat\'ern covariance function which is more general than the Gaussian covariance function used here, but the simulation on functional regression model does not show any improvement. On another direction, one can use Gaussian process models with a non-zero mean. As indicated in \cite{shi07,shi08}, this mean structure can be beneficial when some extra scalar predictors exist. We have also tried this but do not see such advantages of modeling mean curve together with the covariance structure in our simulation.

\begin{table}
\caption{Simulation results for data generated from the Gaussian process model.\label{tab:gp}}
\vspace{0.1in}
\centering
{
\tiny{\begin{tabular}{cccccccc}
\hline\hline
& $s^2$ & $\tau^2$ & $\rho_1$ & $\rho_2$ & MSE & Coverage & Length   \\
\hline
full model & 2.399& 0.051& 20.96& 0.203 & 0.8360& 95.46\%& 3.655\\
           &(1.88,3.08)&(0.046,0.055)&(18.6,23.1)&(0.191,0.215)&&&\\
$m=30, q=10$ &&&&&&&\\

predictive process &2.642&0.051&21.44&0.204& 0.8374&33.35\%& 0.476\\
                   &(2.03,3.49)&(0.046,0.055)&(19.42,23.57)&(0.194,0.215)&&&\\
modification 1 &&&&&&95.78\%& 3.756\\

modification 2 &2.233&0.050&20.22&0.205&0.8356&95.63\%&3.659\\
	       &(1.713,3.005)&(0.046,0.054)&(17.86,23.09)&(0.197,0.216)&&&\\
$m=20, q=40$ &&&&&&&\\

predictive process &4.830&0.298&28.78&0.232&1.008&42.91\%&0.750\\
                   &(3.39,6.72)&(0.273,0.325)&(27.18,30.21)&(0.203,0.252)&&&\\
modification 1 &&&&&&96.60\%&4.400\\

modification 2 &2.621&0.053&23.52&0.205&1.052&94.30\%&4.612\\
               &(2.090,3.264)&(0.048,0.058)&(21.26,27.12)&(0.194,0.216)&&&\\
\hline
\end{tabular}}}
\end{table}

%
%
%
%

\begin{table}
\caption{Simulation results for data generated from the functional regression model.\label{tab:fda}}
\vspace{0.1in}
\centering
{
\tiny{\begin{tabular}{cccccccc}
\hline\hline
& $s^2$ & $\tau^2$ & $\rho_1$ & $\rho_2$ & MSE & Coverage & Length   \\
\hline
full model &9.310& 0.195& 44.82& 0.362 & 1.055& 95.63\%& 2.868\\
           &(5.15,20.00)&(0.17,0.21)&(37.02,60.40)&(0.319,0.406)&&&\\
$m=30, q=10$ &&&&&&&\\

predictive process &9.675&0.198&42.59&0.353& 1.109&50.64\%& 0.596\\
                   &(4.85,18.94)&(0.179,0.219)&(35.25,55.69)&(0.291,0.401)&&&\\
modification 1 &&&&&&94.96\%& 2.800\\

modification 2 &9.718&0.196&42.31&0.356&1.116&95.04\%&2.832\\
	       &(5.28,20.93)&(0.18,0.21)&(37.32,49.58)&(0.319,0.405)&&&\\
$m=20, q=40$ &&&&&&&\\

predictive process &33.031&0.282&86.16&0.360&1.146&40.44\%&0.605\\
                   &(11.19,75.69)&(0.252,0.303)&(77.86,97.32)&(0.336,0.385)&&&\\
modification 1 &&&&&&96.64\%&3.370\\

modification 2 &14.974&0.197&68.03&0.334&1.155&94.59\%&3.128\\
               &(9.24,24.03)&(0.175,0.221)&(57.20,80.87)&(0.299,0.372)&&&\\
\hline
\end{tabular}}}
\end{table}

\begin{table}
\caption{Comparison of prediction errors using different functional regression methods.\label{tab:comp}}
\vspace{0.1in}
\centering
{
{\begin{tabular}{cccccc}
\hline\hline
LININT & LININT2 & LINPT & LINPT2 & KERNEL & OKERNEL   \\
\hline
3.571&0.376&3.742&3.988&2.070&1.993\\

\hline
\end{tabular}}}
\end{table}

\begin{table}
\caption{Prediction errors for the four weather stations using the Gaussian process model, the predictive process approximation, and the linear integral model (\ref{eqn:integral}). \label{tab:weather}}
\vspace{0.1in}
\centering
{
{\begin{tabular}{ccccc}
\hline\hline
&Montreal &Edmonton & Pr. Ruppert & Resolute    \\
\hline
Gaussian process &1.22&0.54&18.85&0.10\\
Predictive process & 1.23 & 0.65 & 19.37&0.11\\
Integral model& 1.10&0.95&31.23&0.18\\
\hline
\end{tabular}}}
\end{table}

\newpage

\bibliographystyle{Chicago}
\bibliography{papers.txt,books.txt}

\end{document}